\documentstyle[]{tonet}
\def\review#1, #2, 1#3#4#5, #6 {{\sl#1\/} {\bf#2} (1#3#4#5) #6 }
\input boxedeps.tex
\SetRokickiEPSFSpecial  
\HideDisplacementBoxes
\def\d{\dagger}
\newcommand{\be}{\begin{equation}}
\newcommand{\eq}{\end{equation}}
\newcommand{\Tr}{{\rm \, Tr}}
\newcommand{\LCTL}{LCTL}
\begin{document}
\hspace*{\fill}FAU-TP3-97-5\\
\hspace*{\fill}hep-th/9707180\\
\markboth{van de Sande and Dalley}{COLOUR DIELECTRIC GAUGE THEORY}
%
\setcounter{part}{10}
%
\title{Colour-Dielectric Gauge Theory on a Transverse Lattice}
%
\author{B. van de Sande\inst{1}, S. Dalley\inst{2}}
%
\institute{\inst{1}Institut F\"ur Theoretische Physik III\\
Staudstra{\ss}e 7, D-91058 Erlangen, Germany
\\
\inst{2} Department of Applied Mathematics and Theoretical Physics\\
Silver Street, Cambridge CB3 9EW, England
}
\maketitle
%
%
%
%
%

We investigate consequences of the effective colour-dielectric
formulation of lattice gauge theory using the light-cone Hamiltonian
formalism with a transverse lattice~\cite{us}.  As a quantitative test
of this approach, we have performed extensive analytic and numerical
calculations for $2+1$-dimensional pure gauge theory in the large $N$
limit.   We study the structure of coupling constant space for
our effective potential by comparing with results available from
conventional Euclidean lattice Monte Carlo simulations of this system.
In particular, we calculate and measure the scaling behaviour of the
entire low-lying glueball spectrum, glueball wavefunctions, string
tension, asymptotic density of states, and deconfining temperature.

The recent Euclidean Lattice Monte Carlo (ELMC) simulations of Teper
\cite{teper} have shown that pure non-Abelian gauge theory behaves
much the same way in three as in four dimensions: a discrete set of
massive boundstates are generated by a linearly confining string-like
force.  Moreover, Teper has performed calculations for $N=2$, $3$, and
$4$, allowing an extrapolation to large $N$.  The
large-$N$ limit is convenient, though not essential, for our
Light-Cone Transverse Lattice (\LCTL{}) formulation.  Our formulation
offers the rare possibility of describing the parton, constituent, and
string behaviour of hadrons in one framework. The relationship between
these pictures, each very different but equally successful, remains
one of the outstanding enigmas of QCD.

We characterise a dielectric formulation as one in which gluon fields,
or rather the $SU(N)$ group elements they generate, are replaced by
collective variables which represent an average over the fluctuations
on short distance scales, represented by complex $N\times N$ matrices $M$.  
These dielectric variables carry colour and
form an effective gauge field theory with classical action minimised
at zero field, meaning that colour flux is expelled from the vacuum at
the classical level.  The price one pays for starting with a simple
vacuum structure is that the effective action will be
largely unknown and must be investigated {\em per se}.

Starting with the Wilson lattice action~\cite{wilson}, we take the
continuum limit in the $x^0$ {\em and} $x^2$ directions, leaving the
transverse direction $x^1$ discrete.  Replacing the link variables
$U\in SU(N)$ with $N \times N$ complex matrics $M$, one derives a
transverse lattice action whose form was first suggested in
Ref.~\cite{bard}
\be 
A = \int dx^0 dx^2 \sum_{x_1} \left( \Tr\left\{
D_{\alpha} M_{x^1}(D^{\alpha} M_{x^1})^{\d}\right\} - {a \over 4g^2}
\Tr\left\{F_{\alpha\beta}F^{\alpha\beta}\right\} - V_{x^1}[M]\right)
\label{lag}
\eq
where $\alpha,\beta \in \{0,2\}$ and
\be
D_{\alpha} M_{x^1}  =  \left(\partial_{\alpha} +i A_{\alpha} (x^1)\right)
        M_{x^1}  -  iM_{x^1} A_{\alpha}(x^1+a)  \; .
\label{covdiv}
\eq
$M_{x^1}$ lies on the link between $x^1$ and $x^1 + a$ while
$A_{\alpha}(x^1)$ is associated with the site $x^1$. $V_{x^1}[M]$ is a
purely transverse gauge invariant effective potential.
Next we introduce light-cone co-ordinates $x^{\pm} = x_{\mp} = (x^0 \pm
x^2)/\sqrt{2}$ and quantise by treating $x^+$ as canonical time.  
The theory has a conserved current
\be
  J^{\alpha}_{x^1}
  =  i \left[M_{x^1} \stackrel{\leftrightarrow}{D^{\alpha}} 
M_{x^1}^{\d}  + M_{x^1-a}^{\d} 
\stackrel{\leftrightarrow}{D^{\alpha}} M_{x^1-a} \right] \label{J}
\eq
at each transverse lattice site $x^1$.  If we pick the light-cone
gauge $ A_- =0$ the non-propagating field $A_+$ satisfies a simple
constraint equation at each transverse site
$\left(\partial_{-}\right)^{2} A_+(x^1) = g^2 J^{+}_{x^1}/a$.  Solving
this constraint leaves an action in terms of the dynamical fields
$M_{x^1}$
\be
A =  \int dx^+ dx^- \sum_{x^1} \Tr \left\{
\partial_{\alpha} M_{x^1} \partial^{\alpha} M_{x^1}^{\dagger} 
+ {g^2 \over 2a} 
J^{+}_{x^1} \frac{1}{\left(\partial_-\right)^2} J^{+}_{x^1} \right\}
- V_{x^1}[M] \; .  \label{action}
\eq
At large $N$, Eguchi-Kawai reduction~\cite{ek} introduces considerable
simplification. For $P^1=0$ the theory is isomorphic to one
compactified on a one-link transverse lattice, {\em id est} we can
simply drop the argument $x^1$ or $l$ from $M$ in all of the previous
expressions. Effectively one is now dealing with a $1+1$-dimensional
gauge theory coupled to a complex scalar field in the adjoint
representation (with self-interactions).

For the transverse effective potential $V[M]$, we will include
all Wilson loops and products of Wilson loops up to fourth order in
link fields $M$:
\begin{eqnarray}
  V[M] & = & \mu^2  \Tr \left\{MM^{\d}\right\} + {\lambda_1 \over a N}
\Tr \left\{MM^{\d}MM^{\d} \right\} \nonumber  \\
&&+ {\lambda_2 \over a N}  \Tr \left\{MM^{\d}M^{\d}M\right\}
+ {\lambda_3 \over a N^2} \Tr \left\{M^{\d}M\right\} 
\Tr \left\{M^{\d}M\right\} 
\label{pot}
\end{eqnarray}
Note that the last term above, which might appear suppressed at
large $N$, is in fact non-zero only for 2 particle Fock states.

Let us introduce creation/annihilation operators
\be
 M_{x^1}(x^-)  =  
\frac{1}{\sqrt{4 \pi }} \int_{0}^{\infty} {dk \over {\sqrt k}}
    \left( a_{-1}(k,x^1) e^{ -i k x^-}  +  \left(a_{+1}(k,x^1)\right)^{\d} 
e^{ i k x^-} \right) \; .
\eq
In the associated Fock space, we include only states annihilated by the 
charge $\int dx^- J^+$.
This gives a Hilbert space formed from 
all possible closed Wilson loops of link
modes  $a_{\pm}$ on the transverse lattice. 
Thus, a typical $p$-link loop will be something like
\be
\Tr  \left\{ a_{+1}^{\d}(k_1) a_{-1}^{\d}(k_2)
a_{-1}^{\d}(k_3)
\cdots a_{+1}^{\d}(k_p)\right\} \,  |0\rangle 
\eq
where the number of $+1$'s equals the number of $-1$'s,
$\sum_{m=1}^{p} k_m = P^+$, and $k_m \ge 0$.  At large $N$ we need
only study the dynamics of single connected Wilson loops in the
Hilbert space since the loop-loop coupling constant is of order $1/N$.
These loops may be thought of as `bare' glueballs, and the problem is
to find the linear combinations that are on mass shell.  Neglecting
$k_m=0$, which is consistent with expanding about the $M=0$ solution
of the dielectric regime, the Fock vacuum is an eigenstate of the full
light-cone Hamiltonian $P^- \left|0\right\rangle = P^+
\left|0\right\rangle =0$.  

The theory possesses several discrete symmetries. Charge conjugation
induces the symmetry ${\cal C}: \, a_{+1,ij}^{\d} \leftrightarrow
a_{-1,ji}^{\d}$.  There are two orthogonal reflection symmetries
${\cal P}_1$ and ${\cal P}_2$ either of which may be used as `parity'.
If ${\cal P}_1: x^1 \to -x^1$, we have ${\cal P}_1: \, a_{+1,ij}^{\d}
\leftrightarrow a_{-1,ij}^{\d}$.  If rotational symmetry has been
restored in the theory, states of spin ${\cal J} \neq 0$ should form
degenerate ${\cal P}_1$ doublets $|+{\cal J}\rangle \pm |-{\cal
J}\rangle$~\cite{teper}.  We use ``spectroscopic notation'' $|{\cal
J}|^{{\cal P}_1 {\cal C}}$ to classify states.

\begin{figure} 
\centering
\BoxedEPSF{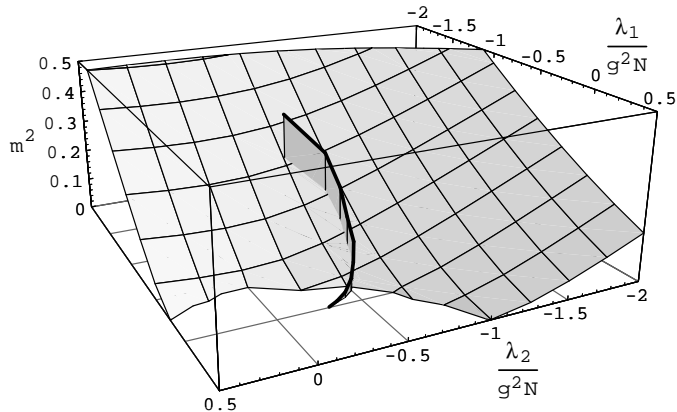 scaled 600}
\caption{ Mass $m^2= \mu^2 a/(g^2 N)$ such that the lowest $M^2$
eigenvalue is zero vs $\lambda_1/(g^2 N)$ and  $\lambda_2/(g^2 N)$
for $p \le 6$, $K=14$, $\lambda_3 =100 g^2 N$.  
Below this surface the spectrum is tachyonic and below $m^2=0$ our
quantisation breaks down.
The dark line is  the scaling trajectory. 
\label{fig3}}
\end{figure}

For $\lambda_1$ and $\lambda_2$ small there is very little mixing
between Fock states of different number of link modes $p$.
In this case a mass eigenstate $|\Psi\rangle$ has predominantly a
fixed $p$, the mass increasing with $p$.  For a given $p$, the energy
also tends to increase with the number of nodes in the wavefunction
$f$ due to the $J\left(\partial_-\right)^{-2}J$ term (\ref{lag}),
which is in fact a positive contribution.  Thus one expects the lowest
two eigenstates to be approximately
\be
\int_{0}^{P^+} dk \, f_{+1,-1}(k,P^{+} - k) \, \Tr 
\left\{ a_{+1}^{\d}(k) a_{-1}^{\d}(P^+-k) \right\} |0\rangle
\eq
with the lowest state having a symmetric wavefunction
$f_{+1,-1}(k,P^+-k)$, corresponding to $0^{++}$, and first excited
state having $f_{+1,-1}$ antisymmetric with one node, corresponding to
$0^{--}$.  The next highest states are either a 4-link state with
positive symmetric wavefunctions $f_{+1,+1,-1,-1}$ and
$f_{+1,-1,+1,-1}$ or a symmetric 2-link state with $f_{+1,-1}$ having
two nodes. In the glueball spectrum we identify the latter states as
$0^{++}_{*}$ and $2^{++}$, respectively, although actual eigenstates
are a mixture of these.

In order to fix the coupling constants in the effective potential, we
perform a least $\chi^2$ fit to Teper's ELMC large $N$ extrapolated
spectrum.  As we shall later show, the mass in units of the coupling
$m^2= \mu^2 a/(g^2 N)$ is a measure of the lattice spacing while the
other terms in our effective potential $\lambda_1$, $\lambda_2$, and
$\lambda_3$ must be found from the fitting procedure.  We will also
determine $g^2 N/a$ based on a fit, which we check self-consistently
with measurements of the string tension.

%
%
%
%
\begin{figure} 
\centering
\BoxedEPSF{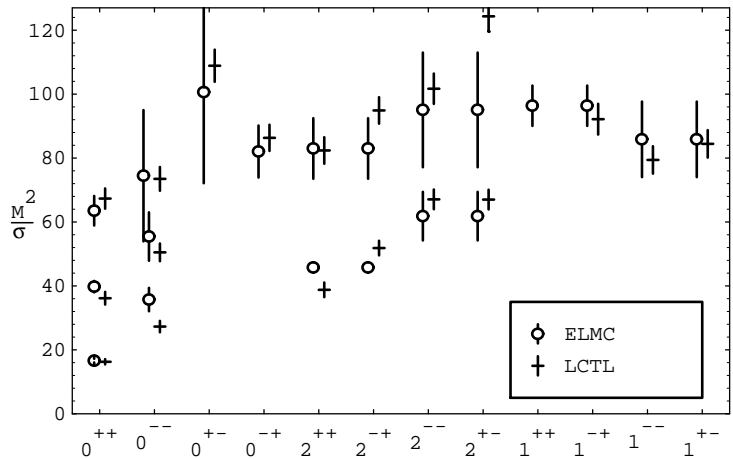 scaled 666}
\caption{
A fit of our \LCTL{} extrapolated results against
Teper's ELMC large $N$ extrapolated spectrum.
The $M^2$ eigenvalues are shown in units of (Teper's) string tension
for various ${\left|{\cal J}\right|}^{{\cal P}_1 {\cal C}}$.
The parameters are $m^2=0.065$, $\lambda_1= -0.202 g^2 N$,
$\lambda_2=-0.55 g^2 N$, $\lambda_3 = 100g^2 N$, and 
$g^2 N = 3.47 a \sigma$ with $\chi^2= 23.5$.
One finds similar spectra along the entire scaling trajectory in
coupling constant space.
\label{fig4}}
\end{figure}

In our numerical solutions we restrict the number of link fields $p$
in our basis states and discretise momenta by demanding
antiperiodicity of the fields in $x^- \to x^-+L$.  In order to
minimise errors associated with these truncations, we take the spectra
for various $K$ and $p$ truncations and extrapolate to the continuum,
$K,p \to \infty$.  Since we cannot measure $|{\cal J}|$ directly, we
only classified states according to ${\cal P}_1$ and ${\cal C}$ during
the fitting process.  We estimate our one-sigma errors from finite $K$
and $p$-truncation to be roughly $0.05 M^2$.  As we fit the various
couplings as a function of $m^2$, we find a narrow strip in parameter
space where we obtain good agreement with the ELMC spectrum; see
Fig.~\ref{fig3}.  As we shall see, moving along this strip
corresponds to changing the lattice spacing $a$. The strip, where
$\chi^2$ has a local minimum, disappears when $m^2$ is sufficiently
large, indicating that for large enough lattice spacing our truncation
of the effective potential is no longer a good approximation.

The numerical bound from absence of tachyons is shown in
Fig.~\ref{fig3} as a zero-mass surface.
As the transverse lattice spacing vanishes the mass gap should vanish
in lattice units. 
The fixed point for this, which we believe lies somewhere at 
negative $m^2$, should lie on the zero-mass surface, but is inaccessible
to us in the dielectric regime $m^2 > 0$.
Nevertheless the scaling trajectory should gradually approach the
zero-mass surface if it is to eventually encounter the fixed point.

In Fig.~\ref{fig4}, we have plotted a typical spectrum along the
scaling trajectory together with the ELMC results.  For graphing
purposes, we assigned $\left|{\cal J}\right|$ to our spectrum based on
a best fit to Teper's results.  
Although the overall fit with the conventional lattice
results is quite good, we see two deficiencies of our spectrum that
cannot be attributed to $K$ or $p$-truncation errors.  First, we see
that the lowest $0^{--}$ state is too low in energy.  Second, we see
that the lowest parity doublet $2^{\pm+}$ is not quite degenerate.  We
believe that these discrepancies must be due to our truncation of the
effective potential.  Finally, we have made no prediction for the
lowest $1^{++}$ state since it lies too high in the $\left|{\cal
J}\right|^{++}$ spectrum.

%
%
\begin{figure} \BoxedEPSF{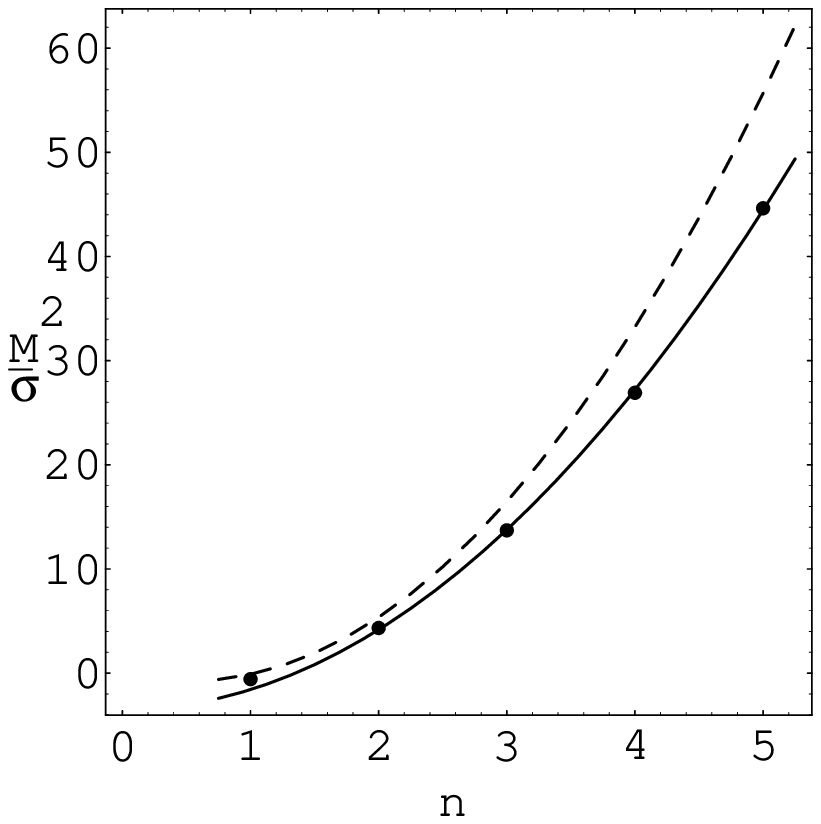 scaled 665}
\hfill\BoxedEPSF{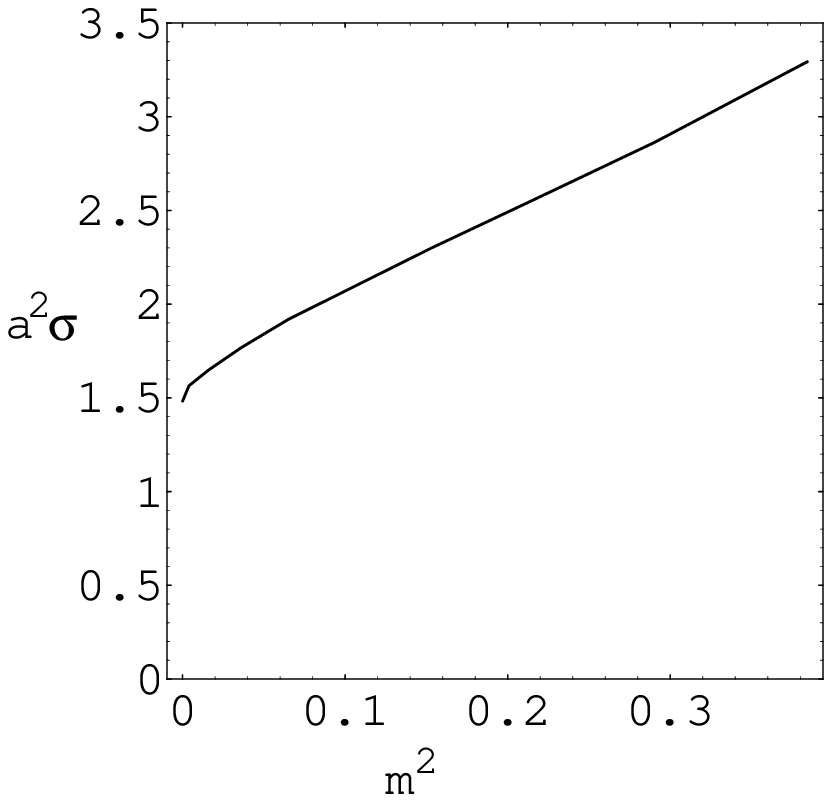 scaled 665}\\
\hspace*{\fill} (a) \hfill \hfill (b) \hfill {}

\caption{(a) Lowest $M^2$ eigenvalue vs 
$n$ for winding modes.
Here, $K=10.5$ or 11, $p \le n+4$, and the couplings 
are taken from Fig.~\protect\ref{fig4}.
Also shown is a numerical fit to $1.92 n^2-3.5$.
The dashed line is from an analytic estimate. 
(b) The lattice spacing in units of the string
tension along the scaling trajectory.  
Here we plot $a^2 \sigma$ vs $m^2$.
\label{fig1}}
\end{figure}

To measure the string tension in the $x^1$ direction (before
Eguchi-Kawai reduction) consider a lattice with $n$ transverse links
and periodic boundary conditions. Constructing a basis of Polyakov
loops, ``winding modes,'' that wind once around this lattice, one may
extract from the lowest eigenvalue $M^2$ vs $n$ the lattice string
tension $a\sigma = \Delta M_n/\Delta n$.  String theory arguments
indicate that oscillations of the winding mode transverse to itself
yield a form $M^2/\sigma = \sigma a^2 n^2 - \pi /3$, the constant
correction being due to the Casimir energy~\cite{luscher}.
Fig.~\ref{fig1} shows a typical $M^2$ vs $n$ plot for winding modes
where we see a good fit to a quadratic.  The constant term -3.5,
however, does not agree well with the expected value of $-\pi/3$.

As a consistency check, we use $\sigma a^2$ from the string tension
measurement together with $g^2 N/(a \sigma)$ from the spectrum to form
the ratio $\sigma g^2$ (as measured by us) to $\sigma g^2$ (as
measured by ELMC).  The result is equal to one $\pm 5\%$ all along the
scaling trajectory.  If we then assume that our $\sigma$ is equal to the
ELMC value of $\sigma$, we can determine the lattice spacing $a$ in
units of the string tension.  
This demonstrates that the mass $m^2=\mu^2 a/(g^2 N)$ determines the
lattice spacing and that the continuum limit occurs at $m^2 <0$; see
Fig.~\ref{fig1}.

%
\begin{figure} 
\centering
\BoxedEPSF{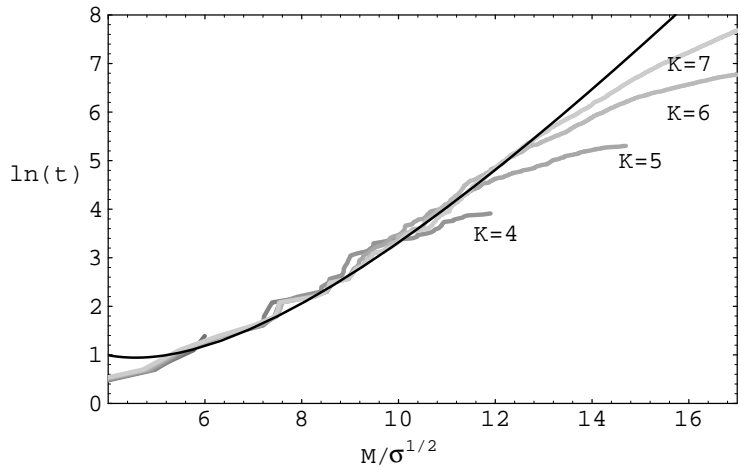 scaled 666}
\caption{
$\ln t$ vs the mass of the $t$-th eigenvalue $M_t$ in
units of the string tension.  Here we have applied 
a cutoff in $K$ only.  Also shown is a least squares fit to 
the $K=7$ data in the range $0.5<\log(t)<5.5$
and couplings from Fig.~\protect\ref{fig4}.
\label{fig21}}
\end{figure}

Another quantity of interest is the deconfinement temperature $T_c$
which may be obtained from the Hagedorn behaviour \cite{hage}
of the asymptotic density of mass eigenstates
$\rho(M) \sim  M^{-\alpha} \exp (M/T_c)$.
The canonical partition function diverges for $T>T_c$. If $\alpha >
(D+1)/2$ it is a phase transition, beyond which the canonical and 
microcanonical ensembles are inequivalent. If $\alpha < (D+1)/2$ the
ensembles are equivalent and $T_c$ represents a limiting temperature
--- the free energy diverges at $T_c$.
%
It is essential to demonstrate that the spectrum is sufficiently
converged in $K$, thus we only fit to the states between 
$0.5<\log(t)<5.5$ for the $K=7$ data in Fig.~\ref{fig21}.  
A numerical fit gives,
%
\be
      \log(t) =3.99 + \frac{M}{0.78\sqrt\sigma} - 
             5.86 \log\left(\frac{M}{\sqrt\sigma}\right) \; .
\eq
Since the density of states is $\rho(M)=dt/dM$, we find $T_c=0.78
\sqrt\sigma$ with an estimated error of at least 10\%.  Due to the
large error, this result is compatible with the Euclidean lattice
result \cite{teper,turin}.  The fact that the power correction $\alpha
\approx 4.86$ in the density $\rho(M)$ is much larger than
$(D+1)/2 = 2$ means we can safely say that the thermodynamic free
energy remains finite at $T_c$, marking a true phase transition
(presumably deconfinement) rather than a limiting temperature.  In
addition, we have an analytic estimate suggesting that $T_c$ lies in
the range $0.81 \sqrt{\sigma} < T_c < 0.98 \sqrt{\sigma}$, also in
agreement with the ELMC result.

%
%
%
%

%
%

\begin{thebibliography}{100}

\bibitem{us} 	S. Dalley and B. van de Sande, Cambridge University report
	No.\ DAMTP-97-16, {\tt hep-ph/9704408}.

\bibitem{teper} M. Teper, \review Phys.\ Lett., 289B, 1992, 115; 
           \review Phys.\ Lett., 311B, 1993, 223; 
 \review Nucl.\ Phys.\ (Proc.\ Suppl.), B53, 1997, 715; 
Oxford University Report OUTP-97-01P (unpublished) {\tt hep-lat/9701004}.

\bibitem{wilson} K. G. Wilson, \review Phys.\ Rev.\ D, 10, 1974, 2445.

\bibitem{bard} W. A. Bardeen and R. B. Pearson, 
             \review Phys.\ Rev.\ D, 14, 1976, 547; 
W. A. Bardeen, R. B. Pearson, and E. Rabinovici, 
           \review Phys.\ Rev.\ D, 21, 1980, 1037.

\bibitem{ek} T. Eguchi and H. Kawai, \review Phys.\ Rev.\ Lett., 48, 1983, 1063.  























\bibitem{luscher} M. L\"uscher, \review Nucl.\ Phys., B180, 1981, 317; 
 P. de Forcrand, H. Schneider, M. Teper, \review Phys.\ Lett., B160, 1985, 137.

\bibitem{hage} R. Hagedorn, \review Nuovo Cimento Suppl., 3, 1965, 147; 
S. Frautchi, \review Phys.\ Rev.\ D, 3, 1971, 2821;  
R. D. Carlitz, \review Phys.\ Rev.\ D, 5, 1972, 3231.






\bibitem{turin} M. Billo, M. Caselle, A. D' Adda, and S. Panzeri,
\review Int.\ J.\ Mod.\ Phys.\ A, 12, 1997, 1783.








%















\end{thebibliography}
\end{document}